\documentclass[10pt,onecolumn,notitlepage]{revtex4-1} 

\usepackage[utf8]{inputenc}
\usepackage[T1]{fontenc}
\usepackage{amsfonts}
\usepackage{amsmath}
\usepackage{braket}
\usepackage{dsfont}
\usepackage{soul}
\usepackage[mathscr]{euscript}
\usepackage{graphicx}
\usepackage{enumitem}
\usepackage{fancyhdr}

\begin{document}

\setlength{\bibsep}{0.0pt}
\bibliographystyle{unsrt}
\rfoot{\thepage/11}
\setlength\itemsep{0em}

\setlength{\belowdisplayskip}{0pt} \setlength{\belowdisplayshortskip}{0pt}
\setlength{\abovedisplayskip}{0pt} \setlength{\abovedisplayshortskip}{0pt}

\title{Solution of Cross-Kerr Interaction Combined with Parametric Amplification}

\author{Sina Khorasani}
\affiliation{Vienna Center for Quantum Science and Technology, University of Vienna, Boltzmanngasse 5, 1090 Vienna, Austria}
\email{sina.khorasani@ieee.org}


\begin{abstract}
We present a full operator approach to treatment of the cross-Kerr interaction combined with parametric amplification. It is shown that this problem can be exactly integrated using the method of higher-order operators. While the initial basis is infinite-dimensional, an orthogonal transformation can reduce the problem exactly into a six-dimensional basis which can be integrated conveniently.
\end{abstract}

\flushbottom
\maketitle

\thispagestyle{empty}

\section*{Introduction}

The cross-Kerr Hamiltonian \cite{Khan,Imoto,Combes,Hadfield} is among one of the mostly used nonlinear quantum interactions between two bosonic fields, which describes a wide range of phenomena. In the case of superconducting circuits, this interaction is of primary importance in modeling nonlinearity of quantum circuits, such as quantum bits and parametric amplifiers. The cross-Kerr interaction also appears in the description of optomechanical systems \cite{OptoCross1,OptoCross2}, photon blockade \cite{PhotonBlockade}, phonon blockade \cite{PhononBlockade}, cross-phase modulation \cite{CrossPhase}, and optical lattices \cite{Zoller}. Usually, one field represents a strong or pump field while the other refers to the weak or probe field. In the context of quantum optomechanics \cite{Khan} the physical nature of these two interacting fields could be quite different, referring to the photons and phonons. When combined with a parametric amplification term, then the total interaction Hamiltonian could be a lot more difficult to solve. So far, no exact solution to this problem has been reported to the best knowledge of the authors.

Here, we demonstrate that the cross-Kerr interaction with parametric amplification could be exactly solvable using the method of higher-order operators \cite{Paper1,Paper2,Paper3,SciRep,Nova,Sideband}, which has evolved out of the rich domain of quadratic optomechanics \cite{Bruschi,Liu,Asjad,Liao,Seok,Zhang,Fan}. This method employs a different basis than the simple bath ladder operators, and quite recently has been independently also reported elsewhere \cite{Liu}. It has been shown that the nonlinear analysis of quantum optomechanics using this algebraic method \cite{SciRep,Nova} may yield quantities such as the coherent phonon population, second-order mechanical side-bands, and corrections to the optical spring effect, as well as nonlinear stability maps. Furthermore, a new type of symmetry breaking named as side-band inequivalence is also found using this algebraic method, which refers to inequal detunings in red- and blue-scattered photons \cite{SciRep,Nova,Sideband}. Furthermore, a preliminary study of photon bunching and anti-bunching statistics applied to the lasing threshold has been carried out using this method \cite{Paper2}, and it has been shown that around the lasing threshold, the cavity population of photons reaches the value of $\sqrt{6}-2$. 

In the context of superconducting quantum circuits, the interaction of two pump-probe microwave fields with the transmon qubits is effectively a cross-Kerr nonlinear interaction \cite{Hoi}, and for all practical reasons it has to be followed immediately by a quantum-limited parametric amplifier stage. This combination leads to a cross-Kerr term with parametric amplification, the solution of which is the purpose of the present study. 

The importance of this contribution is two-fold. On the one hand, one may obtain the time evolution of the number of quanta in time. This enables accurate modeling of quantum non-demolition measurements for photons \cite{Hadfield} and phonons \cite{QND}, for instance, where the number of quanta is measured indirectly through an interaction of cross-Kerr type. Secondly, when optomechanical systems are being considered and the nature of the two interacting bosonic baths are different, the noise spectral density is the actual measurable quantity, the estimation of which is discussed here.

Further contributions of this paper are connected to the treatment of multiplicative noise terms, which normally arise in the method of higher-order operators. It has been demonstrated that for the purpose of calculation of the noise spectral density, these can be exactly simplified to a great extent, where the multiplicative operators can be conveniently replaced by their silent or noiseless non-operator parts. 

\section*{Results}
Consider the cross-Kerr interaction \cite{Khan,Imoto,Hadfield} with parametric amplification \cite{Roy,Holmes,Yamamoto}, defined as 
\begin{align}
\mathbb{H}=\hbar\omega\hat{a}^\dagger\hat{a}+\hbar\Omega\hat{b}^\dagger\hat{b}+\hbar g\hat{a}^\dagger\hat{a}\hat{b}^\dagger\hat{b}+\hbar f\left(\hat{b}^2+\hat{b}^{\dagger 2}\right).
\end{align}
Here, $\omega$ and $\Omega$ respectively refer to the pump and probe frequencies with annihilators denoted by $\hat{a}$ and $\hat{b}$, $g$ represents the cross-Kerr nonlinear interaction rate, and $f$ is the parametric amplification rate. Such type of cross-Kerr mixing can happen in a non-linear cavity where the strength of nonlinear interaction is proportional to the energies (or number of quanta) in each of the two fields. The parametric amplification is normally needed for the readout of probe, which is nonlinearly mixed with the pump and then undergoes amplification prior to detection. Obviously, the cross-Kerr interaction allows some information from the number of quanta of the pump field $\hat{a}^\dagger\hat{a}$ to be passed on to the probe field $\hat{b}^\dagger\hat{b}$ during the detection.

This Hamiltonian is usually analyzed using the basis $\{A\}^\text{T}=\{\hat{n},\hat{m},\hat{C},\hat{S} \}$ where $\hat{m}=\hat{b}^\dagger\hat{b}$, $\hat{n}=\hat{a}^\dagger\hat{a}$, and 
\begin{align}
\hat{C}&=\frac{1}{2}\left[(\hat{n}+1)^{-\frac{1}{2}}\hat{a}+\hat{a}^\dagger (\hat{n}+1)^{-\frac{1}{2}}\right], \\ \nonumber
\hat{S}&=\frac{1}{2i}\left[(\hat{n}+1)^{-\frac{1}{2}}\hat{a}-\hat{a}^\dagger (\hat{n}+1)^{-\frac{1}{2}}\right],
\end{align}
are the quadrature operators, satisfying the commutators $[\hat{n},\hat{C}]=-i\hat{S}$, $[\hat{n},\hat{S}]=i\hat{C}$, and $[\hat{C},\hat{S}]=\frac{1}{2}i(\hat{n}+2)^{-1}$. Usage of these quadrature operators might be advantageous in studying some cases, but construction of Langevin equations would require further approximation since these do not form a closed Lie algebra. As a result, their usage normally needs further linearization procedures which as a result deviates from the mathematically exact solution. In our analysis, however, we use a different basis with closed Lie algebra, which not only admits exact solution, but also allows inclusion of a parametric amplification to either of the $\hat{a}$ or $\hat{b}$ fields.

In the present formulation, we exclude the drive term from the Hamiltonian, and instead feed it through the input noise terms to the system. In particular, when the input terms also fluctuate around a non-zero input or drive term, this approach is accurate. Besides simplicity and the rather convenience involved, the other reason is that the drive term normally contains the simple ladder operator such as $\hat{a}$, whose presence changes the operator basis significantly. Any method to circumvent this difficulty could be much helpful in mathematical description of the problem. Furthermore, this picture where noise and drive terms are fed through the same channel to the system is physically consistent and correct.

\subsection*{Langevin Equations}
We try to analyze this type of interaction in an open-system using Langevin equations \cite{Aspelmeyer,Bowen,Meystre,Gardiner1,Gardiner2}
\begin{align}
\frac{d}{dt}\hat{x}&=-\frac{i}{\hbar}[\hat{x},\mathbb{H}]-\sum_j[\hat{x},\hat{a}_j^\dagger]\left(\frac{1}{2}\kappa_j\hat{a}_j+\sqrt{\kappa_j}\hat{a}_{j,\text{in}}\right)+\sum_j\left(\frac{1}{2}\kappa_j\hat{a}_j^\dagger+\sqrt{\kappa_j}\hat{a}_{j,\text{in}}^\dagger\right)[\hat{x},\hat{a}_j],
\end{align}
where $j$ denotes the bosonic bath, $\hat{a}_j$ is the corresponding annihilator, and $\kappa_j$ is the associated coupling/loss rate. Hence, choosing $j=a,b$ implies $\hat{a}_a=\hat{a}$ and $\hat{a}_b=\hat{b}$, and also $\kappa_a=\kappa$ and $\kappa_b=\Gamma$, respectively corresponding to pump and probe, strong and weak fields, or photons and phonons, depending on the nature of the system under study. Furthermore, $\hat{a}_{j,\text{in}}$ is the input quantum noise from the bosonic bath $j$, and $\hat{x}$ is any operator in the system.

Choosing the infinite dimensional closed Lie algebra of higher-order operators
\begin{align}
\{A\}^\text{T}=\{\hat{m},\hat{d},\hat{d}^\dagger,\hat{n}\hat{m},\hat{n}\hat{d},\hat{n}\hat{d}^\dagger,\dots,\hat{n}^j\hat{m},\hat{n}^j\hat{d},\hat{n}^j\hat{d}^\dagger,\dots \},
\end{align}
where $\hat{d}=\frac{1}{2}\hat{b}^2$, with $[\hat{d},\hat{m}]=2\hat{d}$, $[\hat{m},\hat{d}^\dagger]=2\hat{d}^\dagger$, $[\hat{d},\hat{b}^\dagger]=\hat{b}$ and $[\hat{d},\hat{d}^\dagger]=\hat{m}+\frac{1}{2}$ \cite{Paper2} allows construction of linear infinite-dimensional Langevin equations, given as
\begin{align}
\frac{1}{2\Omega}\frac{d}{dt}\left\{A\right\}=\left(i\left[\textbf{M}\right]-\left[\Gamma\right]\right)\left\{A\right\}-i\frac{\alpha}{2}\left\{A_\text{c}\right\}-\left[\sqrt{\Gamma}\right]\left\{A_\text{in}\right\},
\end{align}
in which $\alpha=f/\Omega$ and 
\begin{align}
\left\{A_\text{c}\right\}&=\left\{A_{\text{c},j};j\in\mathscr{N}\right\}=\{0,1,-1,0,\hat{n},-\hat{n},\dots,0,\hat{n}^l,-\hat{n}^l,\dots \},
\end{align}
and
\begin{align}
\left[\Gamma\right]&=\left[\Gamma_j\delta_{ij};i,j\in\mathscr{N}\right] =\text{diag}\{\gamma_1,\gamma_1,\gamma_1,\dots,\gamma_l,\gamma_l,\gamma_l,\dots\},
\end{align}
is a diagonal matrix of normalized loss rates with $\gamma_l=\left[\Gamma+(l-1)\kappa\right]/2\Omega$. Furthermore, the noise input vector is
\begin{align}
&\{A_\text{in}\}^\text{T}=\{\hat{A}_{\text{in},j};j\in\mathscr{N}\}^\text{T}=\frac{1}{\sqrt{2\Omega}}\{\hat{m}_\text{in},\hat{d}_\text{in},\hat{d}^\dagger_\text{in},\dots,(\hat{n}^j\hat{m})_\text{in},(\hat{n}^j\hat{d})_\text{in},(\hat{n}^j\hat{d}^\dagger)_\text{in},\dots \}, 
\end{align}
in which the combined noise terms are constructed following \S{S1} of Supplementary Information according to 
\begin{align}
\sqrt{\gamma_{j+1}}(\hat{n}^j\hat{d})_\text{in}&=\sqrt{j\frac{\kappa}{2\Omega}}\hat{n}^j_\text{in}\hat{d}+\sqrt{\frac{\Gamma}{2\Omega}}\hat{n}^j\hat{d}_\text{in}; j\in\mathscr{Z}^+, \\ \nonumber
\sqrt{\gamma_{j+1}}(\hat{n}^j\hat{d}^\dagger)_\text{in}&=\sqrt{j\frac{\kappa}{2\Omega}}\hat{n}^j_\text{in}\hat{d}^\dagger+\sqrt{\frac{\Gamma}{2\Omega}}\hat{n}^j\hat{d}^\dagger_\text{in}; j\in\mathscr{Z}^+, \\ \nonumber
\sqrt{\gamma_{j+1}}(\hat{n}^j\hat{m})_\text{in}&=\sqrt{j\frac{\kappa}{2\Omega}}\hat{n}^j_\text{in}\hat{m}+\sqrt{\frac{\Gamma}{2\Omega}}\hat{n}^j\hat{m}_\text{in}; j\in\mathscr{Z}^+. 
\end{align}
The single terms are given as
\begin{align}
\hat{n}^{j+1}_\text{in}&=\sqrt{j+1}\left(\hat{n}^j\hat{a}^\dagger\hat{a}_\text{in}+\hat{a}_\text{in}^\dagger\hat{a}\hat{n}^j\right); j\in\mathscr{Z}^+, \\ \nonumber
\hat{m}_\text{in}&=\hat{b}^\dagger\hat{b}_\text{in}+\hat{b}_\text{in}^\dagger\hat{b}, \\ \nonumber
\hat{d}_\text{in}&=\frac{1}{2}\hat{b}\hat{b}_\text{in}+\frac{1}{2}\hat{b}_\text{in}\hat{b}.
\end{align}
At this point, there are three very important facts to take notice of: 
\begin{enumerate}[topsep=0pt,itemsep=0pt,partopsep=0pt,parsep=0pt]
	\item 
	Firstly, the contributing part of the multiplicative operators which operate on the white Gaussian noise processes $\hat{a}_\text{in}$ and $\hat{b}_\text{in}$ as shown in \S{S2} of Supplementary Information are actually the silent or noiseless parts of these operators, which can be found by solving the corresponding Langevin equations with all zero-mean stochastic processes dropped and only keeping the drive terms. The calculation of silent terms will thus be no longer an operator problem, and can be addressed by any appropriate analytical or numerical approach.
	\item
	Secondly, the order of multiplicative terms, as whether they appear on the left or right of the noise terms is found to be immaterial within the accuracy of Langevin equations. This latter and rather important conclusion can be drawn from the last equation which silent operators actually commute with any Gaussian White noise process, following the construction procedure discussed in \S{S1} of Supplementary Information, and is furthermore compatible with the commutation of multiplicative terms with noise operators. 
	\item
	The third issue is connected to the Hermitian conjugates of noise processes, such as $\hat{a}^\dagger_\text{in}(t)$ as opposed to $\hat{a}_\text{in}(t)$. In the frequency domain these are time-reversed conjugates of each other, which happen to be identical by the general laws of the expectation values of Gaussian noise, given by \cite{Gardiner1,Gardiner2} $\braket{\hat{a}^\dagger_\text{in}(w)\hat{a}_\text{in}(W)}=\delta(w+W)$ and $\braket{\hat{a}_\text{in}(w)\hat{a}_\text{in}(W)}=0$. Therefore, while the spectral densities of $\hat{a}_\text{in}(w)$ and $\hat{a}_\text{in}^\dagger(w)$ are evidently equal, they share the same Fourier transform, too. As a result, the Hermitian conjugate can be arbitrarily dropped from or added to the Gaussian White noise processes as long as the noise spectral density is going to be the quantity to be calculated.
\end{enumerate}

Hence, for the purpose of calculation of noise spectral density at non-zero frequencies, the replacements $\hat{d}_\text{in}=b\hat{b}_\text{in}$, and similarly $\hat{n}^{j+1}_\text{in}=(a^\ast +a)n^j\hat{a}_\text{in}$ and $\hat{m}_\text{in}=(b^\ast+b)\hat{b}_\text{in}$ are admissible, where all multiplicative operators can effectively be replaced with their silent contributions. Knowledge of these expressions is extremely helpful in any computation of noise spectral density, especially in the context of the method higher-order operators, where occurrence of multiplicative noise terms is inevitable. 

Also, the dimensionless coefficients matrix $\left[\textbf{M}\right]$ may be decomposed into real-valued $3\times 3$ partitions as
\begin{align}
\left[\textbf{M}\right]=\left[\begin{array}{c|c|c|c|c|c}
\textbf{A} & \textbf{B} & \textbf{0} & \textbf{0} & \textbf{0} & \dots \\ \hline
\textbf{0} & \textbf{A} & \textbf{B} & \textbf{0} & \textbf{0} & \dots \\ \hline
\textbf{0} & \textbf{0} & \textbf{A} & \textbf{B} & \textbf{0} & \dots \\ \hline
\textbf{0} & \textbf{0} & \textbf{0} & \textbf{A} & \textbf{B} & \dots \\ \hline
\vdots & \vdots & \vdots & \vdots & \vdots & \ddots 
\end{array}\right],
\end{align}
in which the partitions are given by
\begin{align}
\left[\textbf{A}\right]=\left[\begin{array}{ccc}
0 & 2\alpha & -2\alpha \\
-\alpha & -1 & 0 \\
\alpha & 0 & 1
\end{array}
\right],
\end{align}
and 
\begin{align}
\left[\textbf{B}\right]=\left[\begin{array}{ccc}
0 & 0 & 0 \\
0 & -\beta & 0 \\
0 & 0 & \beta
\end{array}
\right],
\end{align}
with $\beta=g/\Omega$. Similarly, the normalized decay matrix $[\Gamma]$ can be written as
\begin{align}
\left[\Gamma\right]=\left[\begin{array}{c|c|c|c|c}
\textbf{G}_1 & \textbf{0} & \textbf{0} & \textbf{0} & \dots \\ \hline
\textbf{0} & \textbf{G}_2 & \textbf{0} & \textbf{0}  & \dots \\ \hline
\textbf{0} & \textbf{0} & \textbf{G}_3 & \textbf{0} & \dots \\ \hline
\vdots & \vdots & \vdots & \vdots & \ddots 
\end{array}\right],
\end{align}
in which the partitions are given by $\textbf{G}_j=\text{diag}\{\gamma_j,\gamma_j,\gamma_j\}$.

\subsection*{Diagonalization}
Here, we can show that there exist $3\times 3$ matrices $[\textbf{U}]$ and $[\textbf{V}]$ in such a way that if the $9\times 9$ unimodular transformation matrix $[\textbf{P}]$ with $|[\textbf{P}]|=1$ is constructed as
\begin{align}
[\textbf{P}]=\left[\begin{array}{c|c|c}
\textbf{I} & \textbf{0} & \textbf{U} \\ \hline
\textbf{0} & \textbf{I} & \textbf{V} \\ \hline
\textbf{0} & \textbf{0} & \textbf{I} 
\end{array}
\right],\\ \nonumber
[\textbf{P}]^{-1}=\left[\begin{array}{c|c|c}
\textbf{I} & \textbf{0} & -\textbf{U} \\ \hline
\textbf{0} & \textbf{I} & -\textbf{V} \\ \hline
\textbf{0} & \textbf{0} & \textbf{I} 
\end{array}
\right],
\end{align}
where single lines separate $3\times 3$ blocks, and 
\begin{align}
[\textbf{Q}]=\left[\begin{array}{c||c|c|c|c|c}
\textbf{P} & \textbf{0} & \textbf{0} & \textbf{0} & \textbf{0} & \dots \\ \hline\hline
\textbf{0} & \textbf{I} & \textbf{0} & \textbf{0} & \textbf{0} & \dots \\ \hline
\textbf{0} & \textbf{0} & \textbf{I} & \textbf{0} & \textbf{0} & \dots \\ \hline
\textbf{0} & \textbf{0} & \textbf{0} & \textbf{I} & \textbf{0} & \dots \\ \hline
\vdots & \vdots & \vdots & \vdots & \vdots & \ddots 
\end{array}\right],
\end{align}
where double lines separate $9\times 9$ blocks, and then
\begin{align}
[\textbf{Q}]^{-1}&\left(i[\textbf{M}]-[\Gamma]\right)[\textbf{Q}]=\left[\begin{array}{c|c|c||c|c|c}
i\textbf{A}-\textbf{G}_1 & i\textbf{B} & \textbf{0} & \textbf{0} & \textbf{0} & \dots \\ \hline
\textbf{0} & i\textbf{A}-\textbf{G}_2 & \textbf{0} & \textbf{0} & \textbf{0} & \dots \\ \hline
\vdots & \vdots & \vdots & \vdots & \vdots & \ddots 
\end{array}\right].
\end{align}
This orthogonal transformation reduces the coefficients matrix $i[\textbf{M}]-[\Gamma]$ in such a way that the Langevin equations for the first six elements of $\{A\}$ are isolated. That therefore will reduce the infinite dimensional problem exactly into a six-dimensional problem in the basis
\begin{align}
\{A_6\}^\text{T}=\{\hat{m},\hat{d},\hat{d}^\dagger,\hat{n}\hat{m},\hat{n}\hat{d},\hat{n}\hat{d}^\dagger\}.
\end{align}
To show the existence of such a transformation, we can evaluate the transformed matrix $[\textbf{R}]=[\textbf{Q}]^\textbf{T}\left(i[\textbf{M}]-[\Gamma]\right)[\textbf{Q}]$ first, and then set the first two rows of the third column of the $3\times 3$ partition blocks to zero. This gives to the set of algebraic equations
\begin{align}
i\left(\textbf{A}\textbf{U}-\textbf{U}\textbf{A}+\textbf{B}\textbf{V}\right)-\textbf{G}_1\textbf{U}+\textbf{U}\textbf{G}_3=\textbf{0},\\ \nonumber
i\left(\textbf{A}\textbf{V}-\textbf{V}\textbf{A}+\textbf{B}\right)-\textbf{G}_2\textbf{V}+\textbf{V}\textbf{G}_3=\textbf{0}.
\end{align}
When expanded, these give rise to a total of $18=2\times 9=2\times 3\times 3$ linear algebraic equations in terms of the elements of $\textbf{U}$ and $\textbf{V}$, which conveniently offers a unique solution for nonzero decay matrix $[\Gamma]$. Explicit expressions are not useful and numerical solution can help if needed. But it is not difficult to calculate $\textbf{V}$ from the second equation. Doing this gives
\begin{align}
\textbf{V}&=\frac{1}{\lambda(4\alpha^2-\lambda^2-1)} \left[
\begin{array}{ccc}
0 & 2\alpha\beta(i-\lambda) & -2\alpha\beta(i+\lambda)\\
-\alpha\beta(i+\lambda) & -i\beta(1+\lambda^2) & 0 \\ 
\alpha\beta(i-\lambda) & 0 & i\beta(1+\lambda^2)
\end{array}
\right],
\end{align}
with $\lambda=\kappa/2\Omega$. However, once it is known that $\textbf{U}$ and $\textbf{V}$ do exist, then it is actually unnecessary to calculate them any longer, since the top left $6\times 6$ block of $\textbf{P}$ is nothing but the identity matrix. That means, very surprisingly, that the truncated system of Langevin equations in terms of the operator basis $\{A_6\}$ as in (18) is already exact. Hence, the $6\times 6$ truncated Langevin equations are actually already exact and integrable for the case of cross-Kerr interaction with parametric amplification.

\subsection*{Classical Pump}

When the pump field $\hat{a}$ is so strong that its quantum nature could be neglected, a more compact representation of the cross-Kerr interaction can be obtained. The same procedure can be exactly applied to the first $3\times 3$ block by solving the equation $i(\textbf{A}\textbf{V}-\textbf{V}\textbf{A}+\textbf{B})+\textbf{G}_1\textbf{V}-\textbf{V}\textbf{G}_2=\textbf{0}$ in terms of the elements of $\textbf{V}$. That will make the truncated $3\times 3$ Langevin equations in terms of the operators $\{A_3\}^\text{T}=\{\hat{m},\hat{d},\hat{d}^\dagger\}$ exact and integrable again. This will lead to the relatively simple expression for the $6\times 6$ unimodular matrix $[\textbf{P}]$ as
\begin{align}
[\textbf{P}]=\left[\begin{array}{c|c}
\textbf{I} & \textbf{V} \\ \hline
\textbf{0} & \textbf{I}  
\end{array}
\right],\\ \nonumber
[\textbf{P}]^{-1}=\left[\begin{array}{c|c}
\textbf{I} &  -\textbf{V} \\ \hline
\textbf{0} & \textbf{I}
\end{array}
\right],
\end{align}
while $\textbf{V}$ is again already known from (19). But this will not pull out any information regarding the second other field expressed by the bosonic population operator $\hat{n}$. In the end, it is appropriate therefore and makes sense to assign $\hat{n}$ to the strong field and $\hat{m}$ to the weak field. Under the circumstances where the strong field could be treated classically, then this $3\times 3$ choice of basis is convenient.

Once the system is made integrable, calculation of Noise Spectral Density and time-evolution of operators becomes straightforward, as discussed in \S{S3} and \S{S4} of Supplementary Information, respectively.

\section*{Discussions}

\subsection*{Steady-State}

Suppose that $\hat{a}$ represents the strong pump field. Then, $\sqrt{\kappa\eta}\braket{\hat{a}_\text{in}}$ is the photon input rate to the cavity, which after normalization corresponds to the input optical power as 
\begin{align}
\xi=\frac{1}{2\Omega}\sqrt{\frac{\kappa\eta }{\hbar \omega}P_\text{op}}.
\end{align}
Here, $\eta$ and $P_\text{op}$ respectively are the coupling efficiency and input optical power. Under steady-state where $d/dt=0$, the operators relax to their mean values. Then one may construct a system of equations in terms of the mean field values $\{\bar{m},\bar{d},\bar{d}^\ast,\overline{nm},\overline{nd},\overline{nd^\ast} \}$. Using the further approximation $\bar{a}=\sqrt{\bar{n}}$, $\overline{nm}\approx\bar{n}\bar{m}$ and $\overline{nd}\approx\bar{n}\bar{d}$, as well as $\braket{\hat{b}_\text{in}}=0$, and after significant but straightforward algebra, one may construct the nonlinearly coupled steady state algebraic equations, which can be then solved to yield
\begin{align}
\bar{m}&=\frac{2\alpha^2}{\left(1+\beta\bar{n}\right)^2+\gamma^2-4\alpha^2}, \\ \nonumber
\bar{d}&=-\frac{i\alpha}{i(1+\beta\bar{n})+\gamma}\left(\bar{m}+\frac{1}{2}\right).
\end{align}
Here, $\gamma=\gamma_1$. The mean value of $\bar{n}$ can be obtained by numerical solution of the implicit equation
\begin{align}
\lambda^2\bar{n}\left(\frac{\bar{m}}{2|\bar{d}|}\right)^2=\xi^2.
\end{align}
The above quintic equation in terms of $\bar{n}$ is nonlinearly linked to the normalized pump $\xi$. Here, $\bar{m}$ and $\bar{d}$ are taken from the previous equations (23). The expression within the parentheses is numerically of the order of $4$ for typical choice of cavity parameters, and the quintic equation conveniently offers only one single positive real root for $\bar{n}$ for most range of the input power. This is while in standard optomechanics, this ratio has been shown to be roughly or extremely close to $2$ for respectively side-band resolved or Doppler cavities.

\subsection*{Variations}

Now that the steady-state equations are known, all operators are replaced by their respective variations around their mean values, and non-zero mean drive and constant terms can be dropped. Doing this, simplifies the problem as the $3\times 3$ set of normalized dimensionless Langevin equations, given by
\begin{align}
\frac{d}{d\tau}\left\{
\begin{array}{c}
\delta\hat{d} \\
\delta\hat{d}^\dagger\\
\delta\hat{m} 
\end{array}
\right\}
&=\left[
\begin{array}{ccc}
-i(1+\beta\bar{n})-\gamma & 0 & -i\alpha \\
0 & i(1+\beta\bar{n})-\gamma & i\alpha \\
2i\alpha & -2i\alpha & -\gamma
\end{array}
\right] \left\{
\begin{array}{c}
\delta\hat{d} \\
\delta\hat{d}^\dagger\\
\delta\hat{m} 
\end{array}
\right\}
-\sqrt{\gamma}\left\{
\begin{array}{c}
\bar{b}\hat{y}_\text{in}\\
\bar{b}^\ast\hat{y}_\text{in}^\dagger\\
\bar{b}\hat{y}_\text{in}^\dagger+\bar{b}^\ast\hat{y}_\text{in}
\end{array}
\right\}.
\end{align}
Here, $\tau=2\Omega t$ is the normalized time, $\hat{y}_\text{in}=\hat{b}_\text{in}/\sqrt{2\Omega}$ is the normalized noise input with the normalized symmetrized spectral density $S_{YY}(w)=\frac{1}{2}$, and $\bar{b}=\sqrt{2\bar{d}}$ is known from solution of (24) and then (23).

We now adopt the definitions
\begin{align}
\{\delta\hat{A}\}^\text{T}&=\{\delta\hat{d},\delta\hat{d}^\dagger,\delta\hat{m} \}, \\ \nonumber
[\textbf{N}]&=\left[
\begin{array}{ccc}
-i(1+\beta\bar{n})-\gamma & 0 & -i\alpha \\
0 & i(1+\beta\bar{n})-\gamma & i\alpha \\
2i\alpha & -2i\alpha & -\gamma
\end{array}
\right], \\ \nonumber
\{\hat{A}_\text{in}\}^\text{T}&=\left\{
\bar{b}\hat{y}_\text{in},
\bar{b}^\ast\hat{y}_\text{in}^\dagger,
\bar{b}\hat{y}_\text{in}^\dagger+\bar{b}^\ast\hat{y}_\text{in}
\right\},
\end{align} 
which allows us to rewrite (25) in the compact form
\begin{align}
\frac{d}{d\tau}\{\delta\hat{A}(\tau)\}=[\textbf{N}]\{\delta\hat{A}(\tau)\}-\sqrt{\gamma}\{\hat{A}_\text{in}(\tau)\}.
\end{align}
These equations can be numerically integrated to study the evolution of number of quanta $\hat{m}(\tau)$, where $\hat{y}_\text{in}$ is the stochastic noise input to the system.

\subsection*{Noise Spectral Density}

Taking the Fourier transform in normalized frequency units of $w=\omega/2\Omega$ gives
\begin{align}
\{\delta\hat{A}(w)\}=\sqrt{\gamma}\left([\textbf{N}]-iw[\textbf{I}]\right)^{-1}\{\hat{A}_\text{in}(w)\}.
\end{align}
Using the input-output relation \cite{Gardiner1,Gardiner2} we have
\begin{align}
\{\hat{A}_\text{out}(w)\}&=\left\{[\textbf{I}]-\gamma\left([\textbf{N}]-iw[\textbf{I}]\right)^{-1}\right\}\{\hat{A}_\text{in}(w)\}=[\textbf{S}(w)]\{\hat{A}_\text{in}(w)\}.
\end{align}
Here, we refer $[\textbf{S}(w)]$ as the scattering matrix. Once $[\textbf{S}(w)]$ is known, we can obtain $S_{DD}(w)$ from \cite{Bowen}
\begin{align}
&S_{DD}(w)=|\bar{b}[S_{11}(w)+S_{13}(w)]+\bar{b}^\ast [S_{12}(w)+S_{13}(w)]|^2 S_{YY}(w).
\end{align}
It is ultimately possible to recover $\bar{S}_{BB}(w)$ from $S_{DD}(w)$, which is the desired measurable spectrum, as shown in \S{S3} of Supplementary Information through the transformation
\begin{align}
S_{BB}(w)&=\frac{1}{2}+\mathscr{F}\left\{\sqrt{\frac{1}{2}\mathscr{F}^{-1}\left\{S_{DD}(w)-\frac{1}{2}\right\}(t)}\right\}(w),
\end{align}
in which $\mathscr{F}$ denotes the Fourier transformation, and based on which we may now define $\bar{S}_{BB}(w)=\frac{1}{2}[S_{BB}(w)+S_{BB}(-w)]$ as the symmetrized noise spectrum.

\subsection*{Reflectivity}
We first notice that the Langevin equation for $\delta\hat{d}$ is independent of $\delta\hat{d}^\dagger$ and vice versa, which greatly simplifies the analysis. However, the same is not true for the scattering matrix $[\textbf{S}(w)]$, whose top-left $2\times 2$ block must be diagonalized first to correctly separate contributions from $\hat{d}$ and $\hat{d}^\dagger$.

Let us assume that $[\Sigma(w)]$ is the reflection scattering matrix defined as
\begin{align}
[\Sigma(w)]=[\textbf{I}]+\gamma\left([\textbf{N}]-iw[\textbf{I}]\right)^{-1}.
\end{align}
This scattering matrix is different from $[\textbf{S}(w)]$ defined in (29), since input shines from the outside whereas for the purpose of noise spectral density calculations, noise is generated from within the cavity. Therefore, defining $R(w)=b_\text{out}(w)/|b_\text{in}(w)|$ with $\phi=\angle R(w)$, we have
\begin{align}
\frac{1}{2}R^2(w)&=\frac{1}{2}\left[\Sigma_{11}(w)e^{2i\phi}+\Sigma_{12}(w)e^{-2i\phi}\right]+\Sigma_{13}(w),\\ \nonumber
\frac{1}{2}R^{\ast 2}(w)&=\frac{1}{2}\left[\Sigma_{21}(w)e^{2i\phi}+\Sigma_{22}(w)e^{-2i\phi}\right]+\Sigma_{23}(w).
\end{align}
These can be solved to find the phase $\phi$ as
\begin{align}
\sin(2\phi)=i\frac{\Sigma_{13}(w)-\Sigma_{23}^\ast(w)}{\Sigma_{11}(w)-\Sigma_{21}^\ast(w)-\Sigma_{12}(w)+\Sigma_{22}^\ast(w)}.
\end{align}
This offers the solution
\begin{align}
e^{\pm 2i\phi}&=\frac{\Sigma_{23}^\ast(w)-\Sigma_{13}(w)}{\Sigma_{11}(w)-\Sigma_{21}^\ast(w)-\Sigma_{12}(w)+\Sigma_{22}^\ast(w)}\mp\sqrt{1+\left[\frac{\Sigma_{23}^\ast(w)-\Sigma_{13}(w)}{\Sigma_{11}(w)-\Sigma_{21}^\ast(w)-\Sigma_{12}(w)+\Sigma_{22}^\ast(w)}\right]^2}.
\end{align}
The reflectivity $\mathcal{R}(w)$ and transmissivity $\mathcal{T}(w)$ now can be easily found from the relationship
\begin{align}
\mathcal{R}(w)&=|R^2(w)|,\\ \nonumber
\mathcal{T}(w)&=1-|R^2(w)|.
\end{align}

\subsection*{Fully Linearized Scheme}

Setting up the fully linearized Langevin equations for (1) in terms of both operators $\hat{a}$ and $\hat{b}$ gives an identical set of equations to that of fully linearized optomechanics. In fact, all nonlinear interaction Hamiltonian between two bosonic operators, such as standard optomechanics, standard and non-standard quadratic optomechanics, and cross-Kerr interaction, take identical set of fully linearized equations. This is a well-known fact in nonlinear quantum mechanics.

Here, we proceed only by linearization of the probe beam $\hat{b}$ and leave the pump $\hat{a}$ out of basis. This will give the set of equations (33), which after some further linearization becomes
\begin{align}
\frac{d}{d\tau}\left\{\begin{array}{c}
\delta\hat{b} \\
\delta\hat{b}^\dagger
\end{array}\right\}&=\frac{1}{2}
\left[\begin{array}{cc}
-i(2\beta\bar{n}+1)-\frac{1}{2}\gamma & -i4\alpha \\
i4\alpha & i(2\beta\bar{n}+1)-\frac{1}{2}\gamma
\end{array}\right] 
\left\{\begin{array}{c}
\delta\hat{b} \\
\delta\hat{b}^\dagger
\end{array}\right\}-\sqrt{\gamma}\left\{\begin{array}{c}
\hat{y}_\text{in}\\
\hat{y}_\text{in}^\dagger
\end{array}
\right\}.
\end{align} 
Quite clearly, there is no way to determine the operator mean field values $\bar{n}$ and $\bar{b}$ from this analysis, since the pump field $\hat{a}_\text{in}$ is absent. Let us for the moment assume that $\bar{n}$ is determined from the same equation as (24) found in the above for the extended higher-order basis.

\subsection*{Example}

We assume $\omega=2\Omega=2\pi\times 2\text{GHz}$, and the quality factors for both modes are set to $100$. We furthermore set the coupling efficiency as $\eta=0.4$, while the cross-Kerr interaction rate is $g=2\pi\times 100\text{kHz}$ and the parametric amplification rate is $f=2\pi\times 50\text{MHz}$. The ratio $f/g$ is swept across various input pump optical powers $P_\text{op}$ from close to zero up to $4\text{fW}$. At microwave frequencies, the input optical power of $P_\text{op}=1\text{fW}$ corresponds to a normalized photon input rate of $\xi=0.0155$.

\begin{figure}[ht!]
	\centering
	\includegraphics[width=6in]{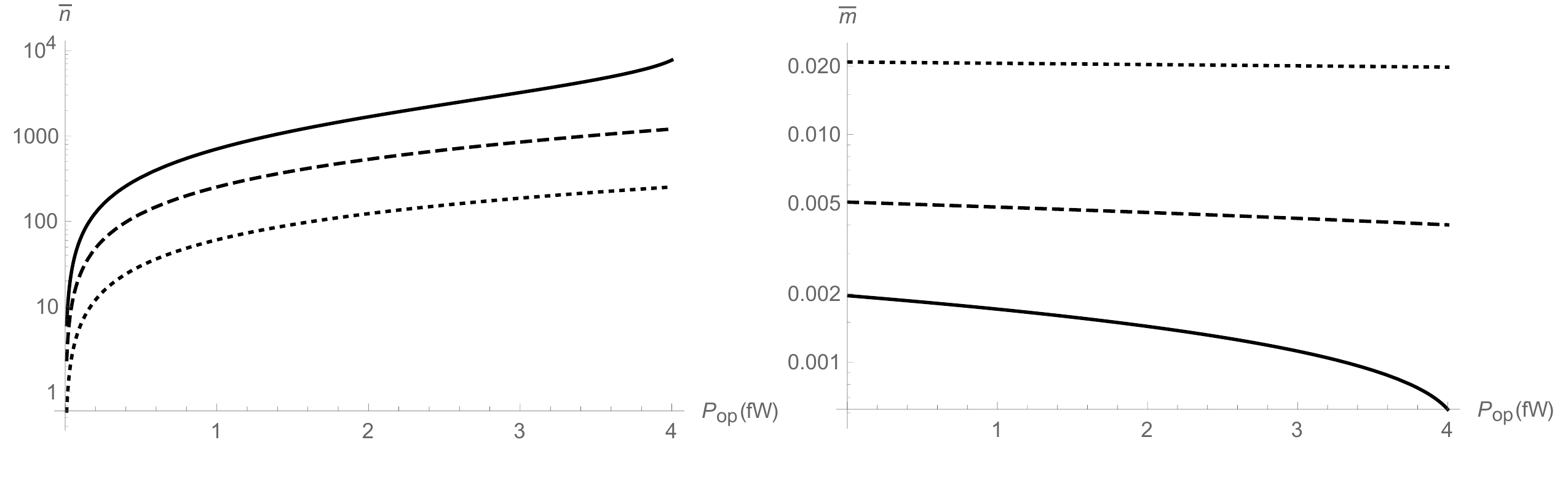}
	
	\caption{Intracavity photon numbers versus input pump power for various ratios of parametric amplification to cross-Kerr interaction rates $\alpha/\beta$ as $\alpha=3.13\times10^2\beta$ (solid); $\alpha=5\times 10^2\beta$ (dashed); $\alpha=10^3\beta$ (dotted): pump photon number $\bar{n}$ (left); probe photon number $\bar{m}$ (right). \label{Fig1}}
\end{figure}

\begin{figure}[ht!]
	\centering
	\includegraphics[width=3in]{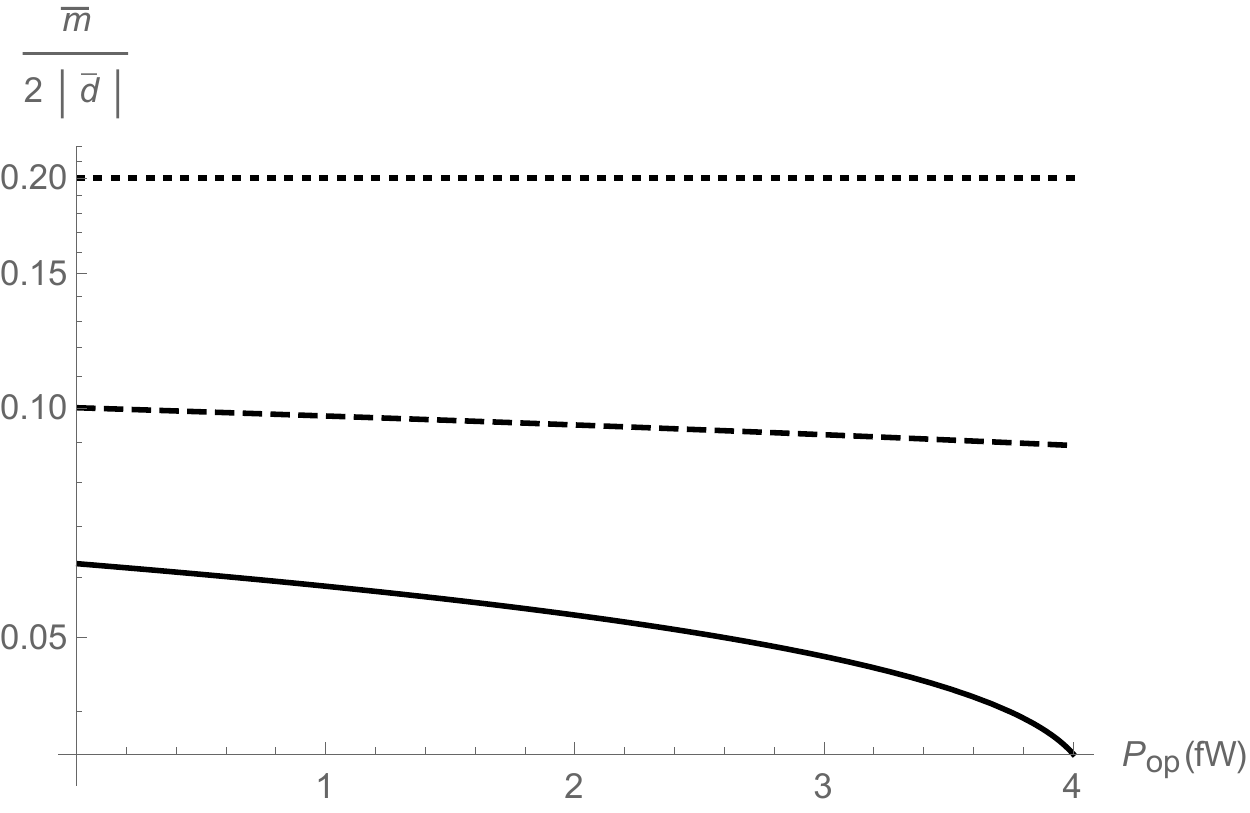}
	\caption{Nonlinearity measure $\bar{m}/2|\bar{d}|$ versus input pump power for various ratios of parametric amplification to cross-Kerr interaction rates $\alpha/\beta$: $\alpha=3.13\times10^2\beta$ (solid); $\alpha=5\times 10^2\beta$ (dashed); $\alpha=10^3\beta$ (dotted). \label{Fig3}}
\end{figure}

We numerically calculate the basic steady-state cavity parameters, including mean pump and probe number of quanta $\bar{n}$ and $\bar{m}$. While $\beta$ is fixed, $\alpha$ is swept over a range of different parameters. As expected, the pump cavity photon number $\bar{n}$ increases nonlinearly with the input power, as shown in Fig. \ref{Fig1}. Meanwhile, the probe cavity photon number $\bar{m}$ is a slowly varying function of pump power, and is instead strongly dependent on the strength of cross-Kerr interaction. Typically, $\bar{m}<1$ and there are normally less than one intracavity probe photons available, as shown in Figs. \ref{Fig1}. The ratio $\bar{m}/2|\bar{d}|$ which describes a measure of the nonlinearity is plotted in Fig. \ref{Fig3}.

\begin{figure}[ht!]
	\centering
	\includegraphics[width=6in]{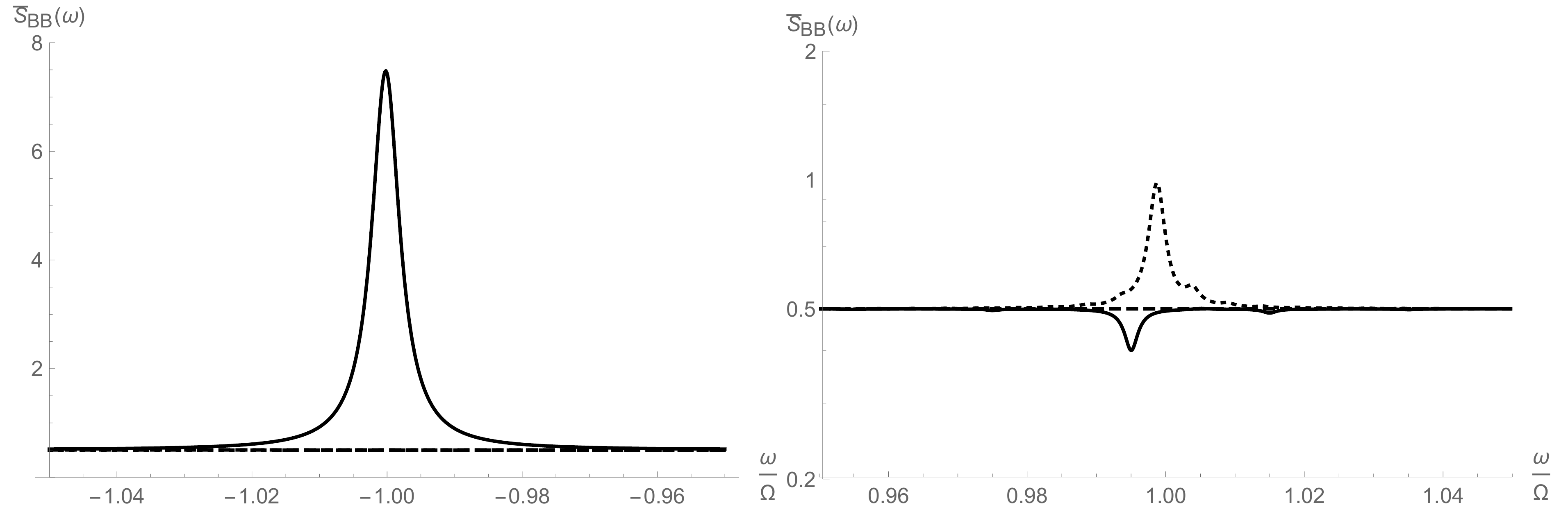}
	\caption{Symmetrized noise spectral density $\bar{S}_{BB}(\omega)=\frac{1}{2}[S_{BB}(\omega)+S_{BB}(-\omega)]$ around major cavity resonance from fully linearized (left) and higher-order (right) operator schemes: $\bar{n}=10^2$ (solid); $\bar{n}=10^3$ (dashed); $\bar{n}=10^4$ (dotted).\label{Fig4}}
\end{figure}

The next three figures illustrate the symmetrized and asymmetric spectral response of the cavity as well as reflectivity around the cavity resonance. The calculations are done for various intracavity pump photon numbers $\bar{n}$, which can be tuned and held constant by fixing the pump power. Since the equation (30) gives the noise spectrum of the higher-order operator $\hat{d}$, we have used (31) to recover the original spectrum of $\hat{b}$. The Fourier and inverse Fourier transforms were taken using discrete Fast Fourier Transform technique with $10^5$ sampling points over the normalized frequency range $[-4,4]$. 

\begin{figure}[ht!]
	\centering
	\includegraphics[width=6in]{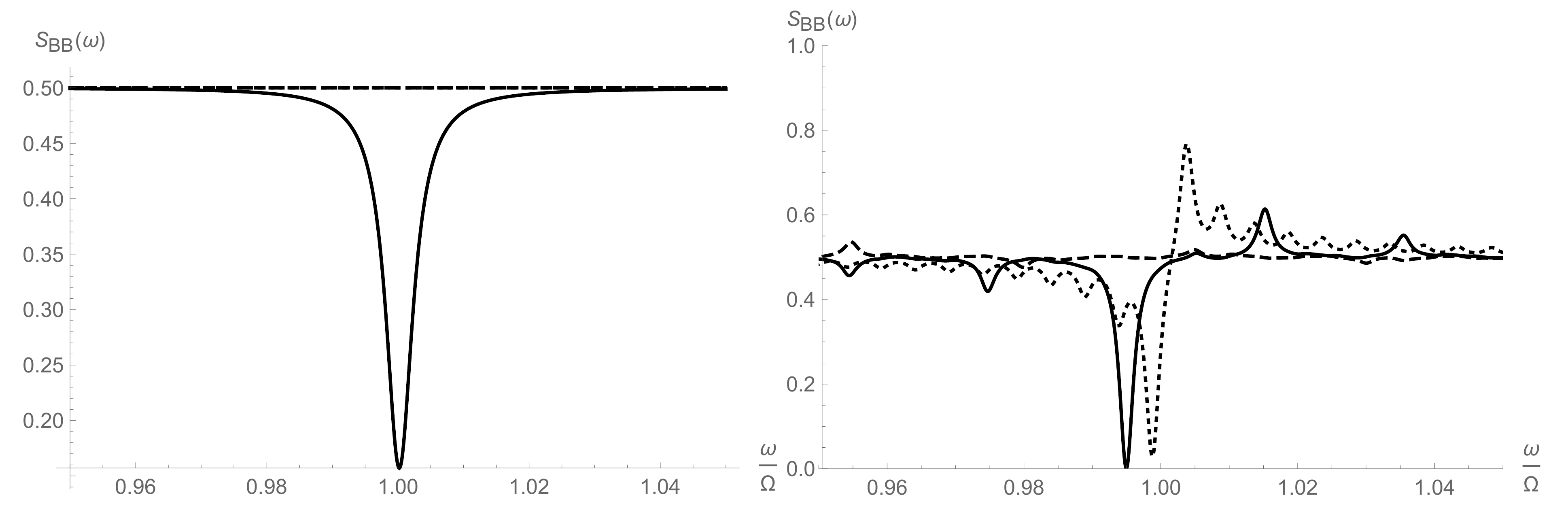}
	\caption{Asymmetric noise spectral density $S_{BB}(\omega)$ around major cavity resonance from fully linearized (left) and higher-order (right) operator schemes: $\bar{n}=10^2$ (solid); $\bar{n}=10^3$ (dashed); $\bar{n}=10^4$ (dotted).\label{Fig5}}
\end{figure}

Not surprisingly, there appears to be some appreciable squeezing around the cavity resonance due to the parameteric amplification, which drives squeeze terms. These are clear from both the symmetrized and asymmetric noise spectra respectively shown in Fig. \ref{Fig4} and Fig. \ref{Fig5}. The squeezing disappears at very large pump drive and is replaced by a peak. At the same time, reflectivity drops around the resonance due to the combined effects of nonlinear cross-Kerr interaction and parametric amplification.

\begin{figure}[ht!]
	\centering
	\includegraphics[width=6in]{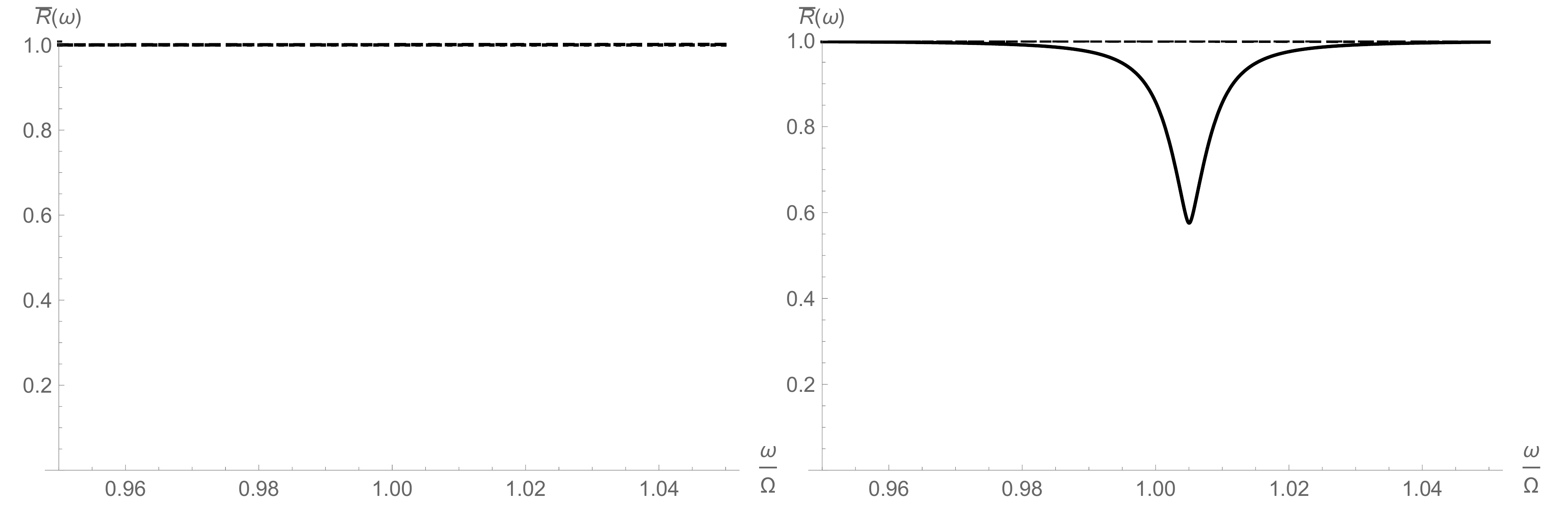}
	\caption{Symmetrized reflectivity of cavity $\bar{\mathcal{R}}(\omega)$ around major cavity resonance from fully linearized (left) and higher-order (right) operator schemes:  $\bar{n}=10^2$ (solid); $\bar{n}=10^3$ (dashed); $\bar{n}=10^4$ (dotted).\label{Fig6}}
\end{figure}

\begin{figure}[ht!]
	\centering
	\includegraphics[width=6in]{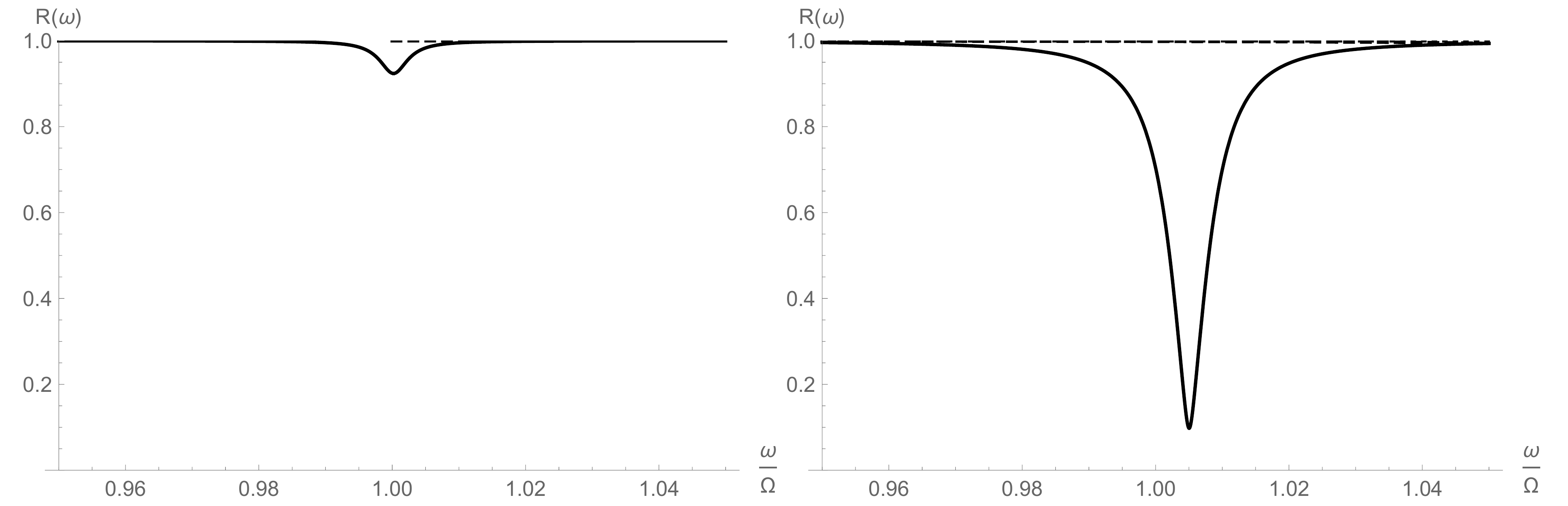}
	\caption{Asymmetric reflectivity of cavity $\mathcal{R}(\omega)$ around major cavity resonance from fully linearized (left) and higher-order (right) operator schemes:  $\bar{n}=10^2$ (solid); $\bar{n}=10^3$ (dashed); $\bar{n}=10^4$ (dotted).\label{Fig7}}
\end{figure}

The reflectivity of the nonlinear cavity is expected to be a function of the pump strength, which has been calculated for both symmetrized and asymmetric forms. These are shown in Figs. \ref{Fig6} and \ref{Fig7} respectively. The maximum reflection dip at resonance for $\bar{n}=10^2$ is well pronounced using the method of higher-order operators.

\section*{Methods}

For an extensive description of theoretical methods, the respectful reader is referred to the Supplementary Information provided along with this article.

\section*{Conclusions}

We presented an exact diagonalization of the cross-Kerr nonlinear interaction with inclusion of parametric amplification. Cases of strong pump and classical pump were considered and also taken into account. It was shown that while it is expected that an infinite-dimensional basis could provide the mathematically exact solution, there exist an orthogonal transformation of infinite order, which can exactly reduce the problem into a finite-order $6\times 6$ formulation. 

\section*{Additional Information}

The author declares no competing interests.

\section*{Acknowledgment}

Preparation of this article would have been impossible without very deep and insightful discussions of this work with Dr. David Edward Brushi at Vienna Center for Quantum Science and Technology and Prof. Andr\'{e} Xuereb at University of Malta. The author would like to hereby extend sincere thanks and humble appreciations for their extremely warm reception and kind support, carefully reading the manuscript, and providing constructive feedback and guidance.  This paper is dedicated to the celebrated artist, Anastasia Huppmann.

\newpage
\rfoot{\thepage/S4}
\section*{ Supplementary Information: Theoretical Methods}
\begin{center}
	\textbf{Solution of Cross-Kerr Interaction Combined with Parametric Amplification}
	
	Sina Khorasani
\end{center}

\setcounter{equation}{0}
\setcounter{page}{1}
\setcounter{table}{0}
\setcounter{section}{0}
\setcounter{figure}{0}
\makeatletter
\renewcommand{\theequation}{S\arabic{equation}}
\renewcommand{\thefigure}{S\arabic{figure}} 
\renewcommand{\thepage}{S\arabic{page}}
\renewcommand{\thetable}{S\arabic{table}}
\renewcommand{\thesection}{S\arabic{section}}
\renewcommand{\bibnumfmt}[1]{[S#1]}
\renewcommand{\citenumfont}[1]{S#1}

\setcounter{equation}{0}
\setcounter{section}{0}

\section{Construction of Noise Terms}\label{Noise}

It is straightforward to see how the noise terms of higher-order operators should be constructed. In the case that the operators are treated fully nonlinearly regardless of their mean values, then the corresponding noise spectral densities could be non-trivial to calculate. This issue, for the case of the squared operator $\hat{d}$ has been discussed in details elsewhere \cite{Paper2S}. But this is appropriate only when the noise spectral density is under consideration. When other quantities are to be measured, for which the non-squared ladder operators might be needed, an iterative approach like $\hat{b}_{j+1}=\hat{d}+\hat{b}_{j}-\frac{1}{2}\hat{b}_{j}^2$ with $\hat{b}_{0}=\hat{1}$ provides uniform convergence to the ladder operator $\hat{b}$.

In order to do this, one may use, for instance, the Langevin equation for $\hat{b}$ and multiply both sides by $\hat{b}$ from left and right. Summing up together, leads to a Langevin equation for $\hat{b}^2$ with a noise input term such as those displayed in (10). The Langevin equation for $\hat{d}$ can be also directly constructed as shown in the article, and that ends up in a noise term as $\hat{d}_\text{in}$ with a decay rate of $\Gamma$. Within the accuracy of Langevin equations, these two noise terms coming from the two approaches should be identical, and this is how one can obtain all the noise terms in (10) in such an iterative manner alike.

One should keep in mind that the Langevin equations are neither exact nor rigorous by nature, as their construction necessitates at least two approximations of non-dispersive coupling and Gaussian white noise. The discussion around this topic is outside of the scope of the present study.

\section{Silence of Multiplicative Operators}\label{Silence}

The first issue to notice in treatment of the multiplicative noise is the dependence of the multiplying operators to the noise terms. These operators also are determined from lower order Langevin equations in which similar noise terms are fed in. Iteratively going back to the lowest order determines that these multiplying operators appear as an infinite series such as 
\begin{equation}
\label{B1}
	\hat{a}(t)=\hat{a}_0(t)+a(t)+a_1(t)\hat{a}_\text{in}(t)+a_2(t)\hat{a}_\text{in}^2(t)+\dots,
\end{equation}
where $\hat{a}_0(t)=\exp\left[( i[\textbf{M}]-[\Gamma] ) t\right]\hat{a}_0(0)$ is the decaying operator term of the homogeneous solution to the system of Langevin equations, which decays to zero, thus taking no part in the steady state solution, and is excluded from contributing to the noise spectral density. Furthermore, $a(t)=a(t)\hat{1}$ is the noiseless or silent part of the operator $\hat{a}(t)$ driven by the external classical pump field, which can in principle be determined from solving the Langevin equations with the stochastic terms dropped, while only keeping the drive terms as input. 

The above term ultimately gets multiplied to another Gaussian white noise term, such as $\hat{a}_\text{in}(t)$ again. Such a multiplicative noise term as $\hat{a}(t)\hat{a}_\text{in}(t)$ will have an expansion given by 
\begin{equation}
\label{B2}
	\hat{a}(t)\hat{a}_\text{in}(t)=a(t)\hat{a}_\text{in}(t)+a_1(t)\hat{a}_\text{in}^2(t)+a_2(t)\hat{a}_\text{in}^3(t)+\dots,
\end{equation}
and so on. It is not difficult to see that as long as a lower-order Gaussian noise term is present, the higher-order terms will have negligible contribution to non-zero absolute (and not detuned) frequencies in the ultimate noise spectral density. This is discussed in \S\ref{HigherOrderNoise}.

In order to establish this, we may define the second-order noise corresponding to the squared process $\hat{c}(t)=\frac{1}{2}\hat{a}^2(t)$, which clearly has a decay rate of $2\kappa$. The corresponding stochastic noise process is
\begin{equation}
\label{B3}
	\hat{c}_\text{in}(t)=\frac{1}{\sqrt{2\kappa}}\hat{a}_\text{in}^2(t).
\end{equation}
The stochastic process $\hat{c}_\text{in}(t)$ is no longer Gaussian white although $\hat{a}_\text{in}(t)$ is a Gaussian white stochastic process by assumption with a symmetrized auto-correlation $\braket{\hat{a}^\dagger(t)\hat{a}(\tau)}_\text{S}=\frac{1}{2}\delta(t-\tau)$. The symmetrized autocorrelation of this higher-order stochastic process in light of the Isserlis-Wick theorem \cite{Paper2S} is thus given by
\begin{align}
\label{B4}
	\braket{\hat{c}_\text{in}^\dagger(t)\hat{c}_\text{in}(\tau)}_\text{S}&=\frac{1}{2\kappa}\braket{\hat{a}_\text{in}^\dagger(t)\hat{a}_\text{in}^\dagger(\tau)}_\text{S}\braket{\hat{a}_\text{in}(t)\hat{a}_\text{in}(\tau)}_\text{S} +2\times\frac{1}{2\kappa}\braket{\hat{a}_\text{in}^\dagger(t)\hat{a}_\text{in}(\tau)}_\text{S}\braket{\hat{a}_\text{in}^\dagger(t)\hat{a}_\text{in}(\tau)}_\text{S}\\ \nonumber
	&=\frac{1}{\kappa}\braket{\hat{a}_\text{in}^\dagger(t)\hat{a}_\text{in}(\tau)}_\text{S}^2=\frac{1}{4\kappa}\delta^2(t-\tau).
\end{align}
The corresponding spectral density of this noise process, being its Fourier transform, simply causes a Dirac delta at zero frequency \cite{Paper2S}. Similarly, all higher-power noise processes will have no contribution to the non-zero frequency of the noise spectral density. As a result, the multiplicative noise (\ref{B2}) can be  effectively truncated as 
\begin{equation}
\label{A5}
	\hat{a}(t)\hat{a}_\text{in}(t)=a(t)\hat{a}_\text{in}(t),
\end{equation}
without causing any error in the non-zero frequencies of the resulting noise spectral density. A more general treatment of the second-order noise processes with Gaussian resonances is discussed elsewhere \cite{Paper2S}.

\section{Noise Spectral Density}\label{SpectralDensity}

Following the general approach to construction of the scattering matrix based on the input-output formalism \cite{Paper2S,Gardiner1S,Gardiner2S}, one may easily show that
\begin{equation}
\label{C1}
	\{\hat{A}_\text{out}(w)\}=\textbf{S}(w)\{\hat{A}_\text{in}(w)\},
\end{equation}
where $\textbf{S}(w)$ is the $6\times 6$ scattering matrix given by
\begin{equation}
\label{C2}
	\textbf{S}(w)=\textbf{I}-[\sqrt{\Gamma}]\left(i[\textbf{M}]-[\Gamma]-iw\textbf{I}\right)^{-1}[\sqrt{\Gamma}].
\end{equation}
Expansion of the output operator array gives
\begin{equation}
\label{C3}
	\hat{A}_{\text{out},j}(w)=\sum_{l=1}^6 S_{jl}(w)\hat{A}_{\text{in},l}(w),
\end{equation}
where $\hat{A}_{\text{in},l}(w)$ are multiplicative noise terms such as $a_l(w)\hat{a}_{\text{in},l}(w)$, where $\hat{a}_{\text{in},l}$ stand for white Guassian White stochastic processes $\hat{a}_{\text{in}}$, $\hat{b}_{\text{in}}$, and their conjugates $\hat{b}_{\text{in}}^\dagger$, $\hat{b}_{\text{in}}^\dagger$, and also $a_l(w)$ are the corresponding Fourier-transformed silent multiplicative terms. This can be correspondingly shown to lead to the noise spectral densities
\begin{equation}
\label{C4}
	S_{AA,j}(w)=\sum_{l=1}^6 \left|\frac{1}{\gamma_l}S_{jl}(w)\ast a_l(w)\right|^2 S_{A_l A_l}(w),
\end{equation}
with the understanding that the terms corresponding to conjugate noise operators are grouped together under the absolute value. Here, $S_{A_l A_l}(w)$ are the symmetrized noise spectral densities of the Gaussian White processes $\hat{a}_{\text{in},l}$. The spectral densities of these processes are typically constants as $\bar{n}_l+\frac{1}{2}$ with $\bar{n}_l$ being thermal occupation number of bosons. For an optical bosonic bath, one may conveniently set $\bar{n}_l=0$, while for phonons $\bar{n}_l$ can be estimated from Bose-Einstein distribution \cite{AspelmeyerS}. Furthermore, the symbol $\ast$ represents convolution in the frequency domain.

The approach provided here, leads to the noise spectral densities of higher-order operators $\hat{d}=\frac{1}{2}\hat{b}^2$, $\hat{d}^\dagger=\frac{1}{2}\hat{b}^{\dagger 2}$ and $\hat{m}$. Neither of these is the directly measurable spectrum, but it is rather the noise spectral density of ladder operator $\hat{b}$ for photons, which can be measured. These necessitates a way to recover the information through what is calculable by the method of higher-order operators.

The symmetrized noise spectral density of $\hat{d}=\frac{1}{2}\hat{b}^2$ is by definition given in terms of the Fourier transform of the corresponding symmetrized auto-correlation function, which is
\begin{align}
\label{C5}
	S_{DD}(w)&=\frac{1}{2\pi}\int_{-\infty}^{+\infty}e^{i w t}\braket{d^\dagger(\tau)d^\dagger(t+\tau)}_\text{S}dw=\frac{1}{8\pi}\int_{-\infty}^{+\infty}e^{i w t}\braket{b^\dagger(\tau)b^\dagger(\tau)b^\dagger(t+\tau)b^\dagger(t+\tau)}_\text{S}dw\\ \nonumber
	&=\frac{1}{4\pi}\int_{-\infty}^{+\infty}e^{i w t}\braket{b^\dagger(\tau)b^\dagger(t+\tau)}_\text{S}^2 dw.
\end{align}
where the last expression is found by application of the Isserlis-Wick theorem and $\braket{b^\dagger(\tau)b^\dagger(\tau)}_\text{S}=0$. By noting the definition of Fourier and inverse Fourier transforms, we get
\begin{align}
\label{C6}
	S_{BB}(w)&=\frac{1}{2}+\mathscr{F}\left\{\sqrt{\frac{1}{2}\mathscr{F}^{-1}\left\{S_{DD}(w)-\frac{1}{2}\right\}(t)}\right\}(w),
\end{align}
where $\frac{1}{2}$ is substrated and added to account for the half a quanta of white noise which is lost in the symmetrization, and if not removed will cause appearance of a non-physical Dirac delta under the square root. While $S_{DD}(w)$ is found from simple scattering matrix calculations, all it takes now to find the measurable quantity $S_{BB}(w)$ is to take an inverse Fourier transform, followed by a square root and another Fourier transform. Similarly, one we may now define $\bar{S}_{BB}(w)=\frac{1}{2}[S_{BB}(w)+S_{BB}(-w)]$ as the symmetrized noise spectrum. The equation (\ref{C6}) is the main key to recover the expected results from the higher-order operator algebra.

\section{Time-evolution of Operators}\label{TimeDomain}

It is easy to obtain the explicit solution to the truncated system of Langevin equations (5)
\begin{align}
\label{D1}
	\left\{A(t)\right\}&=e^{i\left[\textbf{N}\right]2\Omega t}\left\{A(0)\right\}-2\Omega e^{i\left[\textbf{N}\right]2\Omega t} \int_0^t e^{-i\left[\textbf{N}\right]2\Omega \tau}\left(i\frac{\alpha}{2}\left\{A_\text{c}\right\}-\left[\sqrt{\Gamma}\right]\left\{A_\text{in}(\tau)\right\}\right)d\tau,
\end{align}
where $\exp(\cdot)$ represents the matrix exponentiation, and $[\textbf{N}]=[\textbf{M}]+i[\Gamma]$.

\section{Non-negative Integer Powers of Noise}\label{HigherOrderNoise}

It is straightforward to see that any term involving a non-negative integer power of a noise such as $\hat{\alpha}_\text{in}^j (t); j\in\mathscr{N},j>1$ where $\hat{\alpha}_\text{in}^j (t)=\kappa^{\frac{1-j}{2}}\hat{a}_\text{in}^j (t)$ has identically zero contribution to the measured noise spectral density. In order to show this, let us the noise assume the normal autocorrelation
\begin{align}
\label{E1}
	\braket{\hat{\alpha}_\text{in}^{\dagger j}(\tau)\hat{\alpha}_\text{in}^j(t)}_\text{S}=\zeta \exp\left[-\pi \zeta^2 (t-\tau)^2\right].
\end{align}
In the limit of $\zeta\rightarrow +\infty$ this will settle back to the expected Dirac delta's function $\delta(t-\tau)$. The autocorrelation of the measurable optical field is connected to the operator $\hat{a}(t)$, which by means of the Isserlis-Wick theorem becomes 
\begin{align}
\label{E2}
	\braket{\hat{a}^{\dagger }(\tau)\hat{a}(t)}_\text{S}&=\left[\frac{\kappa^{j-1}}{j}\braket{\hat{a}_\text{in}^{\dagger j}(\tau)\hat{a}_\text{in}^j(t)}_\text{S}\right]^\frac{1}{j}=\left(\frac{\kappa^{j-1}\zeta}{j}\right)^{\frac{1}{j}}\exp\left[-\frac{\pi \zeta^2}{j} (t-\tau)^2\right].
\end{align}
The corresponding noise spectral density in frequency domain, where $w$ is the absolute optical frequency (and not the detuning referenced to a certain non-zero resonance frequency), is given by
\begin{align}
\label{E3}
	\mathscr{F}&\left\{\braket{\hat{a}^{\dagger }(\tau)\hat{a}(t)}_\text{S}\right\}(w)=\left(\frac{\kappa^{j-1}\zeta}{j}\right)^{\frac{1}{j}} \mathscr{F}\left\{\exp\left[-\frac{\pi \zeta^2}{j} (t-\tau)^2\right]\right\}(w)|_{\tau=0}
	=\frac{\sqrt{j}}{j^\frac{1}{j}}\left(\frac{\kappa}{\zeta}\right)^\frac{j-1}{j}\exp\left[-\frac{j w^2}{4\pi \zeta^2}\right],
\end{align}
In the limit of $\zeta\rightarrow +\infty$ with $j>1$ the above expression is identically zero, and hence meeting the claim.

It is equally straightbackward to show that for any white Gaussian noise such as $\hat{a}_\text{in}$ satisfying $\braket{\hat{a}_\text{in}^\dagger(t)\hat{a}_\text{in}(\tau)}_\text{S}=\frac{1}{2}\delta(t-\tau)$, the higher-power noise processes  $\hat{\alpha}_\text{in}^j (t)=\kappa^{\frac{1-j}{2}}\hat{a}_\text{in}^j (t)$ contribute only to the zero frequency of the noise spectral density. To show this, we assume
\begin{align}
\label{E4}
	\braket{\hat{a}_\text{in}^{\dagger}(\tau)\hat{a}_\text{in}(t)}_\text{S}=\frac{\zeta}{2} \exp\left[-\pi \zeta^2 (t-\tau)^2\right],
\end{align}
which again in the limit of $\zeta\rightarrow+\infty$ reproduces the Dirac's delta $\delta(t-\tau)$. Then the Isserlis-Wick theorem for such a Gaussian noise process could be exactly used to write
\begin{align}
\label{E5}
	\braket{\hat{\alpha}_\text{in}^{\dagger j}(\tau)\hat{\alpha}_\text{in}^j(t)}_\text{S}&=\frac{j}{2^j \kappa^{j-1}}\braket{\hat{a}_\text{in}^{\dagger}(\tau)\hat{a}_\text{in}(t)}_\text{S}^j=\frac{j\zeta^j}{2^j \kappa^{j-1}}\exp\left[-\pi j \zeta^2 (t-\tau)^2\right].
\end{align}
Taking the Fourier transform from both sides gives the resulting noise spectral density
\begin{align}
\label{E6}
	\mathscr{F}\left\{\braket{\hat{\alpha}_\text{in}^{\dagger j}(\tau)\hat{\alpha}_\text{in}^j(t)}_\text{S}\right\}(w)|_{\tau=0}=\frac{\sqrt{j}\zeta^{j-1}}{2^j \kappa^{j-1}}\exp\left[-\frac{w^2}{4\pi j \zeta^2}\right],
\end{align}
which in the limit of $\kappa=\zeta\rightarrow+\infty$ yields an upper bound to a constant number of quanta $\sqrt{j}/2^j$, being less than $\frac{1}{2}$ for $j>1$. This maximum bound to the background number of added noise quanta due to higher-power noise rapidly decays to zero with increasing $j$.

These limits are physically meaningful as long as cavity linewidth is much larger than the pump laser linewidth, which is quite accurately met in practice. So, when no squeezing is taking place and cavity resonances exhibit noise spectra corresponding to a much larger number of quanta than $\frac{1}{2}$, it should be safe to ignore the effect of square noise terms and higher powers. 

Ultimately, a numerical integration carried out on a nonlinear differential equation with exaggerated noise input amplitude could very visibly distinguish the zero contribution of the higher-power noise terms, quite expectedly, confirming the general above conclusions.

\section{Multiplicative Noise}

It is possible to make simple estimates for $b(w)$ to be used instead of $\bar{b}$ in (30), where a convolution such as (\ref{C4}) would have been needed instead. This can be done by setting up the Langevin equations for operators $\{\hat{b},\hat{b}^\dagger\}$, which are coupled. Then these have to be Fourier-transformed and diagonalized to find $\hat{b}(w)$ explicitly, and the expectation of this expression $b(w)=\braket{\hat{b}(w)}$ can be now used as \cite{SciRepS}
\begin{align}
\label{F1}
	S_{DD}(w)&=\frac{1}{\gamma^2}|b(w)\ast[S_{11}(w)+S_{13}(w)]+b^\ast(w)\ast [S_{12}(w)+S_{13}(w)]|^2 S_{YY}(w).
\end{align}
This process is a lot more complicated and typically can be simplified by direct utilization of (30). Nevertheless, the corresponding Langevin equations are
\begin{align}
\label{F2}
	\frac{d}{d\tau}\left\{\begin{array}{c}
		\hat{b} \\
		\hat{b}^\dagger
	\end{array}\right\}&=\frac{1}{2}
	\left[\begin{array}{cc}
		-i(2\beta\hat{n}+1)-\frac{1}{2}\gamma & -i4\alpha \\
		i4\alpha & i(2\beta\hat{n}+1)-\frac{1}{2}\gamma
	\end{array}\right]
	\left\{\begin{array}{c}
		\hat{b} \\
		\hat{b}^\dagger
	\end{array}\right\}-\sqrt{\gamma}\left\{\begin{array}{c}
		\sqrt{\gamma}\bar{b}+\hat{y}_\text{in}\\
		\sqrt{\gamma}\bar{b}^\ast+\hat{y}_\text{in}^\dagger
	\end{array}
	\right\}.
\end{align}
Replacing $\hat{n}$ with $\bar{n}$, taking the expectation and some simplification gives
\begin{align}
\label{F3}
	\left\{\begin{array}{c}
		b(w) \\
		b^\ast(w)
	\end{array}\right\}&=\gamma\left([\textbf{W}]-iw[\textbf{I}]\right)^{-1}\left\{\begin{array}{c}
		\bar{b} \\
		\bar{b}^\ast
	\end{array}\right\},\\
	[\textbf{W}]&=\frac{1}{2}
	\left[\begin{array}{cc}
		-i(2\beta\bar{n}+1)-\frac{1}{2}\gamma & -i4\alpha \\
		i4\alpha & i(2\beta\bar{n}+1)-\frac{1}{2}\gamma
	\end{array}\right]. \nonumber
\end{align}
Here, the factor $\frac{1}{2}$ is included to take care of normalization with respect to $2\Omega$ rather than $\Omega$. Functions $b(w)$ and $b^\ast(w)$ from solution of (\ref{F3}) can be plugged in the convolutions of (\ref{F1}). Note the cancellation of $\gamma$ as it explicitly appears in (\ref{F1}) and (\ref{F3}).

\end{document}